\documentclass[showpacs,preprintnumbers,amsmath,prl,
graphicx,amssymb,superscriptaddress,twocolumn,nofootinbib]{revtex4}
\usepackage{graphicx}
\usepackage{dcolumn}
\usepackage{bm}

\renewcommand{\today}{\number\day\space\ifcase\month\or January\or 
 February\or March\or April\or May\or June\or July\or August\or 
 September\or October\or November\or December\fi\space\number\year}

\begin{document}
\title{A comparison between the low-energy spectra from CoGeNT and CDMS}

\newcommand{\efi}{Enrico Fermi Institute, Kavli Institute for 
Cosmological Physics and Department of Physics,
University of Chicago, Chicago, IL 60637}
%%%%%%%%%%%%

\affiliation{\efi}
														
\author{J.I.~Collar}\affiliation{\efi}

\begin{abstract}

A side-to-side comparison is established between the nuclear 
recoil energy spectrum from CDMS germanium bolometers and its 
low-energy equivalent for events in the inner bulk volume of a 
CoGeNT germanium diode. Acknowledging the orthogonality of the
background cuts possible with each type of detector and following 
an examination of the uncertainties in these searches, a suggestive 
agreement between these spectra is observed. 

\end{abstract}
%% ENTER PACS NEXT
%\pacs{85.30.-z, 95.35.+d, 95.55.Vj, 14.80.Mz}

\maketitle

%%%%%%%%%%%%%%%%%%%%%%%%%%%%%%%%%%%%%%%%%%%%%%%%%%%%%%%%%%%%%%%%%%%%%%

The CoGeNT collaboration has reported an excess of low-energy events in a P-type 
Point-Contact germanium diode (PPC) operated at the Soudan Underground 
Laboratory \cite{cogent2}, a site in common to the CDMS dark matter search. PPCs are 
particularly sensitive to light-mass (few GeV/c$^{2}$) Weakly Interacting Massive 
Particle (WIMP) dark matter candidates by virtue of their modest electronic noise. 
The observed excess remains after the application of a new background reduction 
technique allowing discrimination against events taking place on PPC surface 
layers \cite{cogent2}.  The possibility of a light-mass 
WIMP being responsible for both the CoGeNT background excess and the annual 
modulation observed by the DAMA/LIBRA experiment \cite{damaclaim} has been considered 
by a number of authors. A viable WIMP phase space (coupling, mass) can be found when, 
for instance, instrumental \cite{danandco} or astrophysical \cite{neal} uncertainties 
are contemplated, or common assumptions on the WIMP-nucleus coupling mechanism are relaxed 
\cite{liam,feng}.  

More recently, the CDMS collaboration performed a search for such light WIMPs in 
existing data from their detectors at Soudan \cite{soudan} and a shallower 
underground site (Stanford Underground Facility, SUF \cite{suf}), 
leading to a claim that exhaustive constraints can be placed on a light-WIMP interpretation for 
the DAMA/LIBRA and CoGeNT anomalies. In this note, several presently unjustified 
choices made to arrive to this claim are examined, concluding that the CDMS low-energy 
recoil-like spectrum is in actuality surprisingly similar to its CoGeNT 
equivalent. This note starts by calling attention to arbitrary choices made in the 
CDMS analysis and background interpretation, and to the uncertainties 
generated by those. It continues with mention of CoGeNT uncertainties, concluding with a comparison 
between the two spectra and a commentary on the possible significance of 
their found similarity. Energy values in this report refer to recoil energy 
unless otherwise stated. This discussion is restricted to CDMS germanium data, 
CDMS silicon data imposing much less restrictive light-WIMP constraints due to 
larger backgrounds \cite{suf}.

The nuclear recoil energy scale for the CDMS germanium detectors at Soudan is 
defined following a less-than-straightforward 
procedure$^{1}$\footnotetext[1]{A discussion of how threshold effects 
can affect this procedure is beyond the scope of this note.} making use of 
reference ionization pulses that are already very noisy below $\sim$10 keV. 
This method is briefly touched upon in \cite{soudan}. An example of the poor quality of these 
ionization pulses, even for the best of all detectors treated, is provided for an event 
at 7.3 keV in Fig.\ 10.6 of \cite{ogburn}. This detector, T1Z5, 
drives the sensitivity of the recent CDMS 
Soudan analysis, displaying the lowest threshold and best 
separation between electron and nuclear recoils  \cite{blas}. These ionization pulses  
disappear into the electronic noise of the ionization channel below 
$\sim$5 keV, i.e., in 
the region of interest (ROI) for light WIMP searches and in particular where a 
meaningful comparison with CoGeNT and DAMA/LIBRA can be established. Not 
being able to exploit reference ionization signals in this region, CDMS authors apply 
an arbitrary (power law) extrapolation of the recoil energy scale below 4 keV, exact 
details of which are not provided in \cite{soudan}. It is hard to imagine that in the 
situation described the recoil energy scale in the ROI could have an entirely 
negligible uncertainty. This last is not claimed in \cite{soudan}, instead the subject goes 
unmentioned, even if it has been recently emphasized within the context of 
the sensitivity of liquid xenon detectors to light-WIMPs \cite{scale}. Ideally, the 
uncertainty in this recoil energy scale should be quantified, with attention paid to 
the effect of any arbitrary choices made (e.g., in the extrapolation) and its effect 
folded into the dark matter sensitivity obtained. If this recoil energy scale 
uncertainty is of order a few percent, as it would be naively estimated, the CDMS 
spectra obtained at the shallow depth of SUF and in Soudan would 
display a significant overlap. An agreement between these spectra would not be 
unexpected, given that for both sites the backgrounds associated to cosmic 
sources are negligible \cite{soudan,suf}.  It is natural to wonder if the 
minor differences between the Soudan and SUF spectra 
(Fig.\ 1) are not just
simply a byproduct of the markedly different methods of energy calibration and data analysis 
employed to derive them. Examples of the uncertainties generated by 
this unwarranted lack of consistency in the treatment of SUF and Soudan CDMS data are 
listed next. 

\begin{figure}
\includegraphics[width=8.5cm]{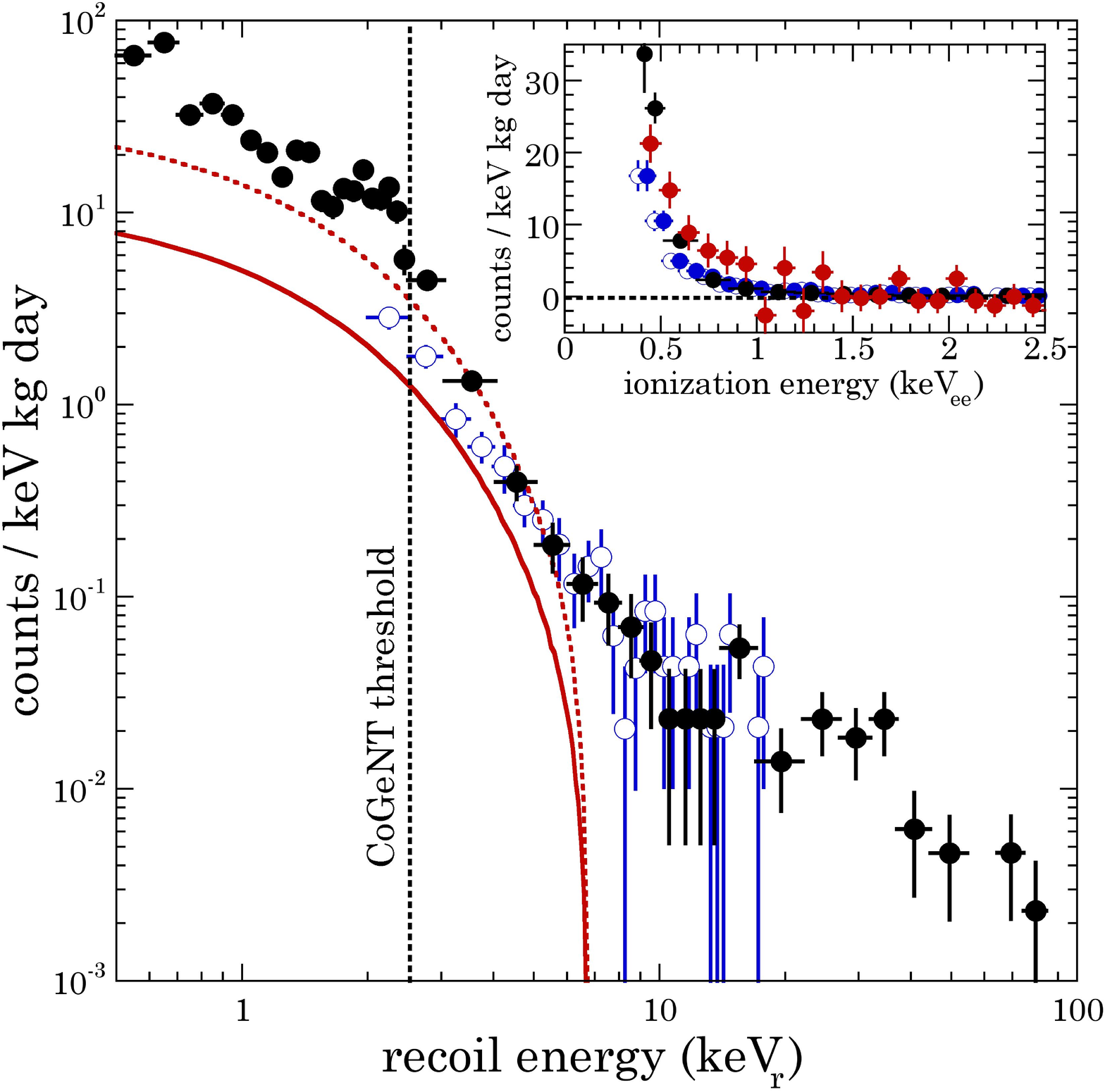}
\caption{CDMS SUF (black dots) and Soudan (blue circles) 
recoil-like spectra, adapted from \protect\cite{soudan,suf}. Red 
dotted line: expectations from a light-WIMP candidate 
(m$_{\chi}\sim$7 
GeV/c$^{2}$, $\sigma_{SI}\sim1.4\times10^{-4}$ pb) able to induce DAMA/LIBRA 
and CoGeNT anomalies \protect\cite{danandco}. Red solid line:  best 
fit to a WIMP of this mass in CoGeNT 
\protect\cite{cogent2}. {\it Inset:} equivalent ionization spectra using a 17\% 
quenching factor \protect\cite{blas}. Red dots: CoGeNT spectrum after 
removal of L-shell EC peak and constant background components 
\protect\cite{cogent2}. Blue dots: Soudan CDMS spectrum 
following the energy shift discussed in the 
text. A new online version of \protect\cite{soudan} acknowledges 
an even 
larger shift ($\times$1.7) at threshold. Vertical error bars are statistical, horizontal 
are energy binning. See text for discussion of other 
uncertainties.}
\end{figure}

A first example is in the choice to employ only the 356 keV $^{133}$Ba calibration 
gamma source peak to define the ionization scale (and in turn the recoil scale) of the 
Soudan CDMS dataset. This approach unrealistically presumes a perfect linearity all 
the way down to a sub-keV ionization threshold. What makes this choice surprising 
is the availability of convenient low-energy ionization energy peaks 
from $^{71}$Ge electron capture (EC) at 10.367 keV and 1.298 keV, following activation during frequent 
neutron calibrations \cite{soudan}. As expected, these peaks were used to establish the 
energy scale of the SUF dataset \cite{suf}. In contrast to this, their use was restricted to a 
cross-check of the assumed linearity of the energy scale in the Soudan analysis. No 
justification for this odd decision is provided in \cite{soudan}. Details have recently 
become available on the quality of this cross-check \cite{blastalk}, indicating that this linearity 
is indeed not optimal at low-energy and possibly rapidly diverging: the peaks appear at 10.34$\pm$0.022 keV and 
1.251$\pm$0.031 keV in the Soudan spectrum. This shift may seem 
modest at $\sim$50 eV ionization energy, but when applied 
to a rapidly increasing near-threshold spectrum (Fig.\ 1, inset) it results in 
an increase in event rate by $\sim$50\%, and a consequent expected relaxation in light-WIMP 
sensitivity for the Soudan dataset by a factor of roughly 
two$^{2}$\footnotetext[2]{As evidenced in Fig.\ 1, the effect of this 
shift should affect any reasonable approach employed to extract 
light-WIMP
limits from the CDMS Soudan dataset. However, the ``optimum 
interval'' method employed in \cite{soudan}, can be particularly 
insensitive to such a small shift, by nature of its definition.}. 
This correction per se would bring the 
Soudan CDMS spectrum to closer agreement with its SUF 
equivalent. Fig.\ 1 (inset) illustrates the magnitude of this correction, 
conservatively assuming an energy-independent constant 50 eV shift. 
The small error bars assigned by CDMS to the position of the  
$^{71}$Ge EC peaks \cite{blastalk} indicate that sufficient statistical information 
existed to attempt a correct low-energy calibration using these  
much more reasonable reference signals.

A second example of lack of consistency between the Soudan and SUF 
CDMS analyses 
is found in the very restrictive acceptance band for nuclear recoils evident in 
Fig.\ 2 of \cite{soudan}. The reader readily notices the differences in this respect with the 
SUF equivalent (Figs.\ 4,9  in \cite{suf}), where the acceptance band is much broader. 
This choice of highly restrictive band seems arbitrary: the logic provided for its 
justification, that it avoids interference from so-called ``zero 
charge'' events, will be 
shown below to be flawed. This change in nuclear-recoil acceptance criteria from 
\cite{suf} to \cite{soudan} leads the reader to wonder about the impact of the 
choice of such cuts on the dark matter limits extracted. An immediate assessment of 
this impact can be found in a recent low-energy analysis of T1Z5 Soudan data in an 
unpublished CDMS thesis \cite{ogburn}, where an irreducible 
background of $\sim$2.3 counts/kg 
day in the recoil energy region 2-5 keV is derived, concluding (after 
examining the  
data from several detectors) that the low-energy spectra collected at 
Soudan and SUF are ``similar''. This clashes with the more recent 
claims in 
\cite{soudan}, even if the 
steps in the analysis of \cite{ogburn} seem to be equivalent. 
The recoil-like background rate that was originally calculated in \cite{ogburn} for T1Z5 represents a 
very significant 70\% increase over 
the later analysis of \cite{soudan}, for the same 2-5 keV energy region, 
identical detector, and 
operation in the same site and 
shielding$^{3}$\footnotetext[3]{This 
increase in rate is a much smaller 3\% for T1Z2, the only other detector in common 
between the analyses in \cite{soudan} and \cite{ogburn}, suggesting that this 
uncertainty stemming from choice of cuts may be exacerbated precisely for the 
dominant T1Z5.}. Such a large dependence on 
the choice
of analysis cuts is not commonplace in dark matter 
searches$^{4}$\footnotetext[4]{Taking CoGeNT as a reference, a 
reanalysis \cite{marino} using simulated pulses 
in lieu of electronic pulser runs to generate surface event 
cuts yields a tight agreement with the 
spectrum in \cite{cogent2}.}, illustrating the fragility of the 
conclusions in \cite{soudan}. This 
subject (effect of choice of analysis cuts on claimed light-WIMP 
sensitivity) is also obviated in \cite{soudan}.

The CDMS collaboration has provided a qualitative description of the backgrounds 
that might be contributing to their low-energy nuclear recoil spectra. This is 
welcome and necessary, and similar to an effort in that same 
direction offered by 
CoGeNT \cite{cogent2}. However, an attempt is also made in 
\cite{soudan,suf} to offer 
quantitative background estimates, claiming that the majority of irreducible recoil-like 
signals stems from so-called ``zero-charge'' events taking place on 
surface layers of the detectors, and that the light-WIMP 
limits obtained are conservative, by not subtracting this background. In order to 
appreciate the difficulty involved in this quantitative exercise, it must be kept in 
mind that on an event-by-event basis and below $\sim$5 keV (higher for detectors other 
than T1Z5), a ``zero charge'' background event and a true neutron-induced low-energy
nuclear recoil are utterly indistinguishable in their characteristics. In other 
words, classifying one as a nuclear recoil and the next as a surface ``zero charge'' 
event would be a fallacy while using present-day CDMS detector technology (at low 
energy, the timing information that allows CDMS detectors to identify surface events 
vanishes with the mentioned wane of ionization pulses). The supposedly dominant ``zero-charge'' 
background is claimed to originate on the subset of surface events 
happening near the edges of the 
cylindrical detector volumes, where charge collection is impeded. To be sure, these 
must be contributing to some unknown extent to the 2-8 keV region, where 
(according to \cite{soudan}) a noticeable WIMP-like rise in the irreducible spectrum is 
observed. The assertion that they constitute the majority of nuclear recoil-like events 
there is nonetheless based on what can be described as an example of circular 
reasoning, detailed next. 

The first step in the quantitative background estimate attempted in 
\cite{soudan} is to fit 
an exponential above 5 keV to the energy spectrum of zero-charge 
events observed during dark matter search runs. A few keV 
above that energy, this family of events becomes readily distinguishable from 
neutron calibration-induced recoils by virtue of their lower ionization yield. CDMS 
authors then proceed to use the low-energy extrapolation of this exponential fit to 
conclude that the majority of recoil-like events under the spectral rise at 2-8 keV 
(Fig.\ 1) can be accounted for by the zero-charge population. This allows them to 
emphasize that their analysis must be conservative, by not attempting to subtract 
this background, producing in the casual reader a false 
impression: the 
problem with this rationale is that at the arbitrarily chosen lower boundary (5 keV) for the 
fitting energy window, bona-fide recoils induced during neutron 
calibrations and zero-charge events already 
share a very significant overlap, i.e., these two populations have 
merged together. This is true even for T1Z5, optimal in 
this respect of 
separation between different families of events and the only detector for 
which the relevant information (ionization yield vs.\ energy) is 
provided in \cite{soudan}. In other words,
true nuclear recoils can be contributing to the exponential-fitted 
region of dark matter 
run spectra, considered as a prior 
in the reasoning above to be exclusively composed  by ``zero charge'' 
events. It is possible to notice this flaw in logic in some of the additional 
figures recently provided in \cite{blastalk}, but not based on the 
information provided in the first online version of 
\cite{soudan}, which did not display neutron calibration data. This is in contrast to the 
more complete disclosure provided for SUF data \cite{suf}. 

What happens if the range of the fitted region is chosen to start somewhere above 5 
keV, where the overlap with nuclear recoils is safely absent? This alternative would 
be considered the correct approach, if one were willing to momentarily ignore the 
strong assumption already made when adopting an arbitrary (exponential) 
behavior to describe the extrapolated energy spectrum of zero-charge events (this 
assumption will be revisited below). Using the data provided in 
\cite{soudan} for T1Z5, 
this reader obtains just a 20\% ``zero charge'' contribution to the 2-8 keV
recoil-like 
spectrum when the lower boundary to the fitting window is minimally 
shifted from 5 keV to 6 keV. This ``zero charge'' contribution is seen to oscillate 
rather wildly with the choice of lower boundary and fitted background model, 
rapidly converging to 100\% as the lower boundary is made any smaller, as expected 
from the flawed reasoning described above. Following 
this criticism, CDMS authors now estimate a 15\% ``zero charge'' 
component with a fitting window starting at 10 keV.

Unfortunately, there are additional inconsistencies in the present attempt to 
quantify low energy backgrounds in the CDMS Soudan 
spectrum$^{5}$\footnotetext[5]{The background budget offered 
in \cite{suf} for SUF 
spectra includes several disconcerting alternative
interpretations, with some proposed 
components being hard to defend. For instance, 10-20\% of the 
low-energy WIMP-like events in SUF are ascribed, without 
justification, to Compton 
scattering of high energy photons. However, the Klein-Nishina 
relation does not provide a mechanism to generate a large low-energy 
excess rise, a point mentioned in the interpretation of CoGeNT 
backgrounds \cite{cogent2}. Similarly, 32\% of SUF WIMP candidates are claimed to originate in the 1.3 keV $^{71}$Ge 
EC line. Their smaller contribution to the ROI is visible as 
a bump around 2 keV (recoil) energy in Fig.\ 1.}. For instance, given that ``zero charge'' 
events are surface events taking place near the edges of the detectors, they would be 
expected to share some characteristics with events identifiable as happening in 
other surface regions of the devices. For example, in their rate per unit surface area. 
This is a natural assumption to make if the dominant source of surface 
contamination is airborne $^{210}$Pb, an origin proposed by CDMS 
authors \cite{bruch}. $^{210}$Pb 
deposition should in principle not affect the edges any more than other surface 
regions during the passive exposure of an unbiased crystal. Another plausible 
characteristic of ``zero charge'' events in common to other surface events might be 
their spectral shape at low-energy, assumed in \cite{soudan} without justification to be 
an exponential. The subject of surface events has been extensively studied by the 
CDMS collaboration elsewhere (e.g., in \cite{bruch}). Examining the region from 2 to 5 keV in 
\cite{soudan}, the irreducible average event rate there 
is $\sim$8 times larger 
than the average surface event rate in the region 10-100 keV 
integrated over the surface area of the detectors \cite{bruch}.  Normalizing 
these rates to per surface area exacerbates this contrast, the edges being only a small 
fraction of the total surface available. It is hard to imagine how such a large 
concentration of surface events on the edges would come to be, especially in the 
airborne contamination interpretation. The discussion in 
\cite{soudan} does 
not attempt to quantify this subject. As for the spectral shape, no rapid rise towards low 
energy can be observed for the surface event selection shown in 
\cite{saab} (which includes detector 
T1Z5), making the choice of an exponential fit to ``zero-charge'' events hard to 
justify. To further illustrate the disarray of background 
interpretations made thus far by 
CDMS authors, it is explicitly mentioned in \cite{ogburn} that ``many of the backgrounds are 
less prominent'' at low energies, and ``beta (surface) contamination almost disappears'', with 
gammas (which can secondarily contribute to surface events via Compton scattering 
\cite{bruch}) ``also decreasing in rate down to a neutron activation line seen in Ge at 1.3 keV, 
below which they are extremely rare''. To summarize, if ``zero charge'' events 
constitute the bulk of the 2-8 keV rise in Fig.\ 1 and they originate on surface events, then they do 
not seem to agree quantitatively (in rate) nor qualitatively (in spectral shape) with 
previous statements made by the CDMS collaboration on the subject of the surface 
contamination of their detectors$^{6}$\footnotetext[6]{Yet another argument against 
the proposed ``zero-charge'' crystal-edge origin for the majority of events 
in the low energy rise in CDMS spectra is the 
common lower value to the recoil-like rate in this energy region 
found across different CDMS germanium crystals: while a few detectors 
may display evidence of surface contamination (in some cases of known 
accidental origin \cite{saab}) 
the majority of the crystals examined in \cite{soudan} exhibit 
narrowly scattered ROI recoil-like rates. This coincidence points at a common source 
affecting all crystals. It would involve a rather precisely-tuned surface 
contamination across detectors that are known to have a different history of 
exposure to $^{210}$Pb \cite{bruch}. More specifically, 
there is no correlation between ROI rate and large differences in $^{210}$Pb contamination 
measured in \cite{bruch} across detector towers.}.

The energy calibration of CoGeNT detectors, single-channel 
devices sensitive to ionization only, is  
straightforward by comparison to CDMS bolometers. It benefits from several narrow peaks of 
cosmogenic origin in the 1-10 keV (ionization) energy range and from 
dedicated measurements of sub-keV quenching factor using a 
monochromatic neutron beam \cite{Ge2}. These devices are however 
subject to their own sources of uncertainty. For instance, in the 
exact fiducial volume following surface-event cuts, presently estimated 
to be $\sim\pm$10\% \cite{cogent2}. A second correction affecting 
CoGeNT is the contribution to the 0.5-0.9 keV 
(ionization) region from L-shell EC of short-lived 
$^{51}$Cr,$^{56}$Ni and $^{56,58}$Co, calculated to be at the level of 
$\sim$15\% for the dataset  
in \cite{cogent2}. This correction can be predicted using the intensity of 
K-shell EC peaks and will be included in an upcoming release of more 
than one year of continuous CoGeNT data-taking. Finally, even if CoGeNT detectors have 
the ability to efficiently reject surface events down to threshold, 
a small fraction can be expected to escape cuts if bulk signal 
acceptance is to remain optimal \cite{cogent2}. A 
planned extension of the CoGeNT target mass to four detectors, each three 
times the mass of the present diode, aims to obtain significantly 
lower background rates and improved threshold and rejection. 

In view of the direct comparison in Fig.\ 1 and the discussion above, 
the conclusions drawn here are diametrically opposite to those in 
\cite{soudan}: once the present uncertainties in CDMS and CoGeNT
are properly accounted for, any significant differences in light WIMP sensitivity 
between these searches
should be traceable to dissimilar statistical 
methods$^{7}$\footnotetext[7]{Contrary to an intent 
clearly expressed in the main text of \cite{suf}, an exclusion plot 
from the CDMS germanium low-energy analysis remains conspicuously absent there. In 
view of Fig.\ 1, any significant exclusion in \cite{suf} of the CoGeNT favored 
light-WIMP parameter 
space should be the result of the statistical 
combination of silicon and germanium limits, both individually unable 
to produce it, or choice of statistical estimator.}. Additionally, 
an inconsistent treatment of the 
CDMS SUF and Soudan datasets contributes to the
differences between their spectra. 
If there is any merit to the open question of a 
possible common cosmological 
origin to the DAMA/LIBRA and CoGeNT anomalies, it would now seem to 
encompass the CDMS recoil-like
spectrum. 

The agreement between CoGeNT and CDMS spectra visible 
in Fig.\ 1 may seem remarkable at first sight. However,
it may very well be the 
result of a simple coincidence. After all, these are contemporary 
technologies using the same target material: it should 
not be surprising that they would display similar 
background-induced limitations to 
their dark matter sensitivity, even if a significant disparity 
in detector internals, handling, activation and shield 
design most certainly leads to different background environments.  
A counterargument to this is the
orthogonality of the background cuts that these technologies have to 
offer. While CDMS can extend a good separation between electron 
recoils and nuclear recoils to their new lower analysis threshold, it 
loses the ability to reject surface events in this energy region. CoGeNT 
provides the converse background-rejection capabilities, 
arguably diminishing the probability of mere chance generating compatible
irreducible spectra. 

This report is based on a presentation at the PCTS Workshop, Princeton 
University, November 2010. Several of the criticisms presented here have been incorporated 
into the latest online version of \cite{soudan}. The author is 
indebted to B.~Cabrera, D.~Hooper, D.~Moore, A.~Robinson and N.~Weiner for  
useful exchanges.

\end{document}